\documentclass[12pt]{amsart}
\usepackage{times}
\usepackage{amssymb, amsmath}
\usepackage{amsthm}
\usepackage{tikz}
\usepackage{bm}
\usetikzlibrary{matrix,arrows}
\usepackage{enumitem}
\usepackage{booktabs}
\usepackage[section]{placeins}

\theoremstyle{plain}

\theoremstyle{remark}

\graphicspath{{./images/}}

\usepackage[a4paper, total={6in, 8in}]{geometry}

\begin{document}

\title[Stock price forecast with deep learning]{Stock price forecast with deep learning}
\author[F. Kamalov, L. Smail, I. Gurrib]{Firuz Kamalov$^1$$^{\boldsymbol{*}}$, Linda Smail$^2$, Ikhlaas Gurrib$^3$}

\address{$^{1}$ Canadian University Dubai, Dubai, UAE.}
\email{\textcolor[rgb]{0.00,0.00,0.84}{firuz@cud.ac.ae}}

\address{$^{2}$ Zayed University, Dubai, UAE}
\email{\textcolor[rgb]{0.00,0.00,0.84}{linda.smail@zu.ac.ae}}

\address{$^{1}$ Canadian University Dubai, Dubai, UAE.}
\email{\textcolor[rgb]{0.00,0.00,0.84}{ikhlaas@cud.ac.ae}}

\date{\today
\newline \indent $^{\boldsymbol{*}}$ Corresponding author
\newline \indent DOI: 10.1109/DASA51403.2020.9317260}

\begin{abstract}
In this paper, we compare various approaches to stock price prediction using neural networks. We analyze the performance fully connected, convolutional, and recurrent architectures in predicting the next day value of S\&P 500 index based on its previous values. We further expand our analysis by including three different optimization techniques: Stochastic Gradient Descent, Root Mean Square Propagation, and Adaptive Moment Estimation. The numerical experiments reveal that a single layer recurrent neural network with RMSprop optimizer produces optimal results with validation and test Mean Absolute Error of 0.0150 and 0.0148 respectively.
\end{abstract}
\maketitle
\section{Introduction}
Predicting the future value of a stock is a notoriously difficult problem. According to the established economic theory - the Efficient Market Hypothesis -  all the publicly available information is already incorporated in the current stock price. Consequently, it should be infeasible to forecast future price without additional information. However, the existing financial markets are not perfectly efficient and it might be possible to discern a hidden pattern using deep learning methods. In this paper, we analyze the performance of various neural network architectures in forecasting the future value of S\&P 500 index.

Our goal is carry out a comparison of fully connected, convolutional, and recurrent neural network models in stock price prediction. Concretely, we apply deep learning techniques to predict the value of S\&P 500 index. The input of the models consists of the index values from the previous 14 days and the output is the next-day value of the index. We include eight different neural network models in our analysis. To expand our analysis we also include three optimization algorithms: Stochastic Gradient Descent (SGD), Root Mean Square Propagation (RMSprop), and Adaptive Moment Estimation (Adam). The choice of optimizers is based on their significance. The SGD, RMSprop, and Adam are the most widely used optimizers in the current literature. Experimenting with various architectures and training parameters we conclude that a simple neural network consisting of a single hidden layer with 6 recurrent nodes achieves a solid performance. We also observe that the Adam optimizer trained with batch size 32 produces good results. The optimal model produces validation and test MAE of 0.0150 and 0.0148 respectively.
We believe the proposed analysis of neural network architectures would be a useful tool for researchers interested in time series forecasting. 

Our paper is organized as follows. In Section 2, we provide a brief discussion of the current literature. In Section 3, we present the proposed network architecture. Section 4 contains the results of the numerical experiments. And Section 5 concludes the paper.

\section{Literature}
Asset price forecasting is a classic problem and there exists a large body of literature devoted to the subject. Traditional approaches employ various technical indicators such as volatility, momentum, and relative strength index \cite{Gurrib3}.
More recently, the popularization of machine learning methods have propelled their application to forecasting. 
The authors in \cite{Borovkova} use an ensemble of Long Short Term Memory (LSTM) models to predict the direction of large cap US stocks. The input data in the model consists of basic price based variables such as Open, Close, High, Close as well as other more advanced technical indicators. The proposed model was found to outperform the benchmark lasso and ridge regression models.
The authors in \cite{Patel} employ various machine learning models including Support Vector Machines (SVM), Artificial Neural Networks (ANN), Random Forest (RF) and naive Bayes classifiers to predict the direction of Indian stock market. Price based indicators are used to create 10 different technical features that are used as inputs to the models. The results indicate that RF outperforms the other tested methods. 
In \cite{Pyo}, the authors apply ANN and SVM to predict the trend of the Korea Stock Price Index. The authors use price based indicators such as moving average as input features for classification. Numerical experiments produce mixed results 
that are not consistent with previous research.
In \cite{Li}, the authors applied extreme machine learning to forecast trading signal in H-share market. The experimental data consisted of the intra-day tick-by-tick data and news archives for their analysis. The results have showed that  the proposed method achieves both high accuracy and fast prediction speed compared to other benchmark methods.

In \cite{Qiu}, the authors proposed a wavelet transform, based on Long short-term memory neural networks (LSTM) and an attention mechanism, to denoise historical stock data, extract and train its features, and establish the prediction model of a stock price. The model was compared with the LSTM model, the LSTM model with wavelet denoising, and the gated recurrent unit NN model on three datasets including the S\&P 500. The proposed model was shown to be significantly better than the other models with a coefficient of determination higher than 0.94. 
To forecast the values of S\&P 500 index, the authors in \cite{Skuratov} used a system that incorporate adaptive filtering, ANNs, and evolutionary optimization. The proposed system was based on ANNs and the empirical mode decomposition (EMD), where the obtained intrinsic mode functions was optimized by genetic algorithms (GA). The proposed system called EMD-GA-ANN was compared to a GA‐ANN trained with a wavelet transform's. The hybrid system outperformed existing predictive systems tested on the same data set. 
The authors in \cite{Lee} proposed a NN using only the data of individual companies to obtain enough data to replace the target stock index data. The short-term stock values were predicted using the trained model together with a sliding window technique. The process considered a heuristic to control the possible extrapolation anomalies of the DNN. The results of experiments showed that the proposed NN model outperformed the model trained directly on the S\&P 500 index data.  


LSTM model was used in \cite{Fischer} to predict directional movements for the constituent stocks of the S\&P 500 market index. The model inputs consisted of sequences of one-day returns of length 240 days. The proposed method is found to outperform random forest, deep neural net, and logistic regression models. The effectiveness of various input variables was investigated in \cite{Liew}. The authors employed Deep Neural Networks (DNNs), RFs, and SVMs to forecast ETF values. It was discovered that volume is an important factor in forecasting. LSTM was also found effective by the authors in \cite{Kamalov3}. The authors showed that LSTM models are significantly more potent than other models in predicting significant changes in stock price. Various attempts have also been made to address the issue of uneven distribution of price changes. Data obtained from a period of economic downturn would contain more negative values than positive which can lead to classification bias \cite{Thabtah}. A common approach to combat imbalanced distribution is through resampling. Alternatively, outlier detection methods can be used in place of the standard classification algorithms \cite{Kamalov2}.

\section{Model specifications}
\begin{figure}[h!]
\centering
\includegraphics[width=0.7\textwidth]{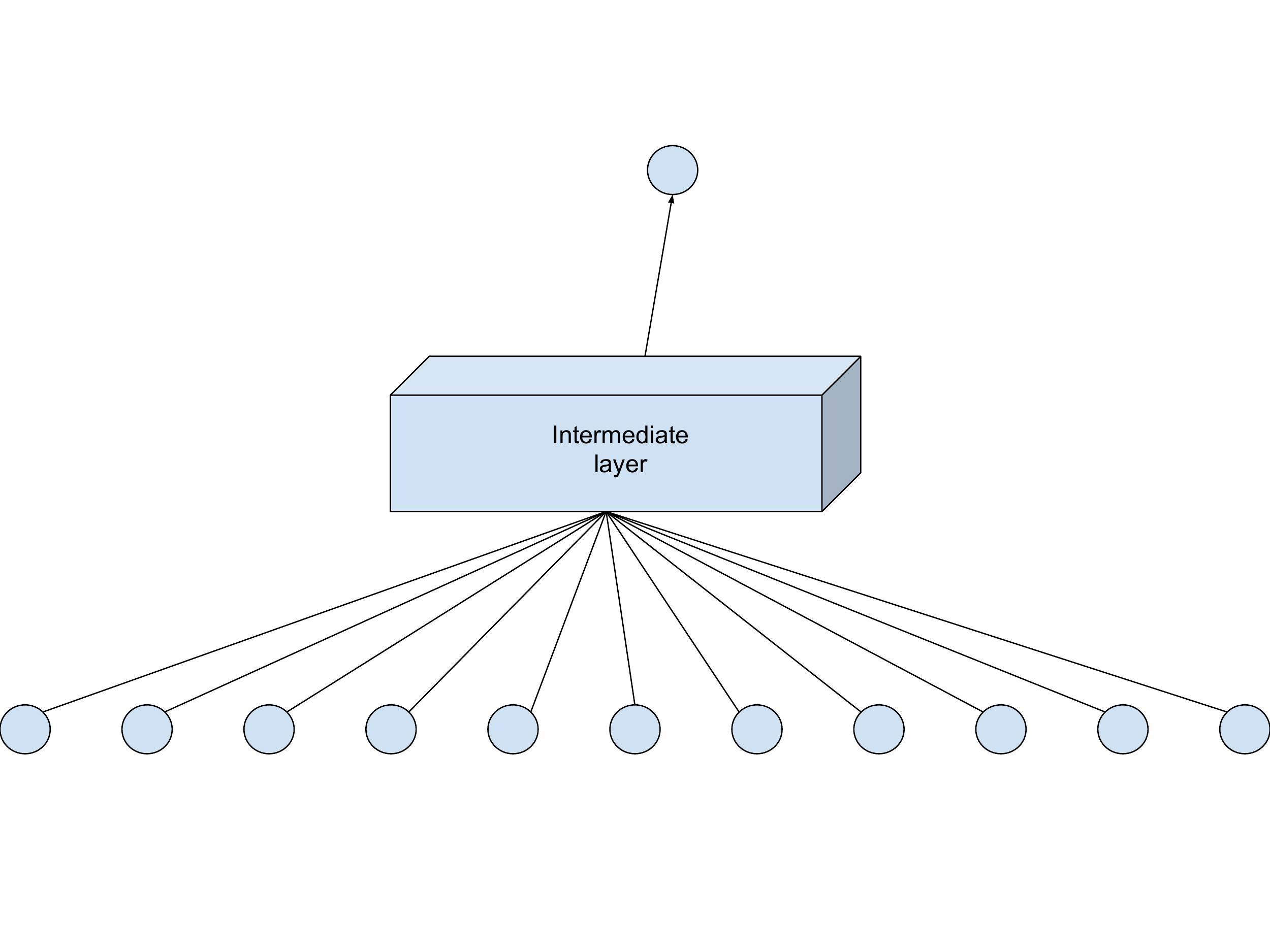}
\caption{The general structure of the neural networks used in the experiments.}
\label{nn_diagram2}
\end{figure}

At the core of our analysis is a collection of neural network models that can be used for prediction. The list is created to include models of different types such as fully connected, recurrent, and convolutional. The general structure of the models is given in Figure~\ref{nn_diagram2}. As shown in the figure, each model consists of a 14-dimensional input layer and a 1-dimensional output layer. The input layer represents index value and volume from the previous 14 trading days while the output layer represents the predicted index value on the next day. Note that 14 is a hyperparameter that is controlled by the user.

We employ eight different architectures for the intermediate layer in Figure~\ref{nn_diagram2}.  The details of the models are presented in Table \ref{tab:models}. In the simplest case, we only use fully connected layers. A single fully connected layer is flexible enough - given enough neurons - to approximate any real valued continuous function on a compact set. However, the underlying assumption when using fully connected layers is that the order of the inputs is irrelevant. Since we are dealing with a time series the order does matter. Hence, we employ recurrent neurons. Recurrent neurons take into account the sequential nature of time series data. As a result, Recurrent Neural Network (RNN) and its variant LSTM are popular architectures in time series analysis. The description of the models in Table \ref{tab:models} follows the standard Keras nomenclature. For instance, the model \texttt{rnn1fc} consists of a recurrent layer with 4 nodes (RNN(4)) followed by a fully connected dense layer with 4 nodes (Dense(4)).

\begin{table}[h!]
\centering
\caption{The descriptions of the models follow Keras nomenclature.}
\label{tab:models}
\begin{tabular}{lll}
\toprule
index & {model} &  description  \\
\midrule
1 & \texttt{fc1} &     Dense(14) \\
2 & \texttt{fc2} &    Dense(14) $\rightarrow$ Dense(7) \\
3 & \texttt{rnn1} &     RNN(4)  \\
4 & \texttt{rnn1fc} &    RNN(4) $\rightarrow$ Dense(4) \\
5 & \texttt{rnn2} &     RNN(6)  \\
6 & \texttt{lstm1} &    LSTM(6) \\
7 & \texttt{conv1} &    Conv1D(4, 3) \\
\bottomrule
\end{tabular}
\end{table}

\subsection{Model training}
The model is trained on daily index values from the past 30 years. The data is divided into train/validation/test sets according to 70/15/15 ratio. We apply the SGD, RMSprop, and Adam optimizers with batch size 32 to train our model. The batch size of 32 is chosen to allow the optimization algorithm to explore a wider range of values in search of the minima. We apply early stopping to avoid model overfitting. Concretely, the validation error is monitored during the training and the process is stopped when the validation error stops decreasing for more than 10 epochs. The shallow architecture of the network serves as another regularization tool - with fewer number of parameters the model is less likely to overfit. The model is implemented using Keras API on Colab platform. 

\begin{figure}[h!]
\centering
\includegraphics[width=0.7\textwidth]{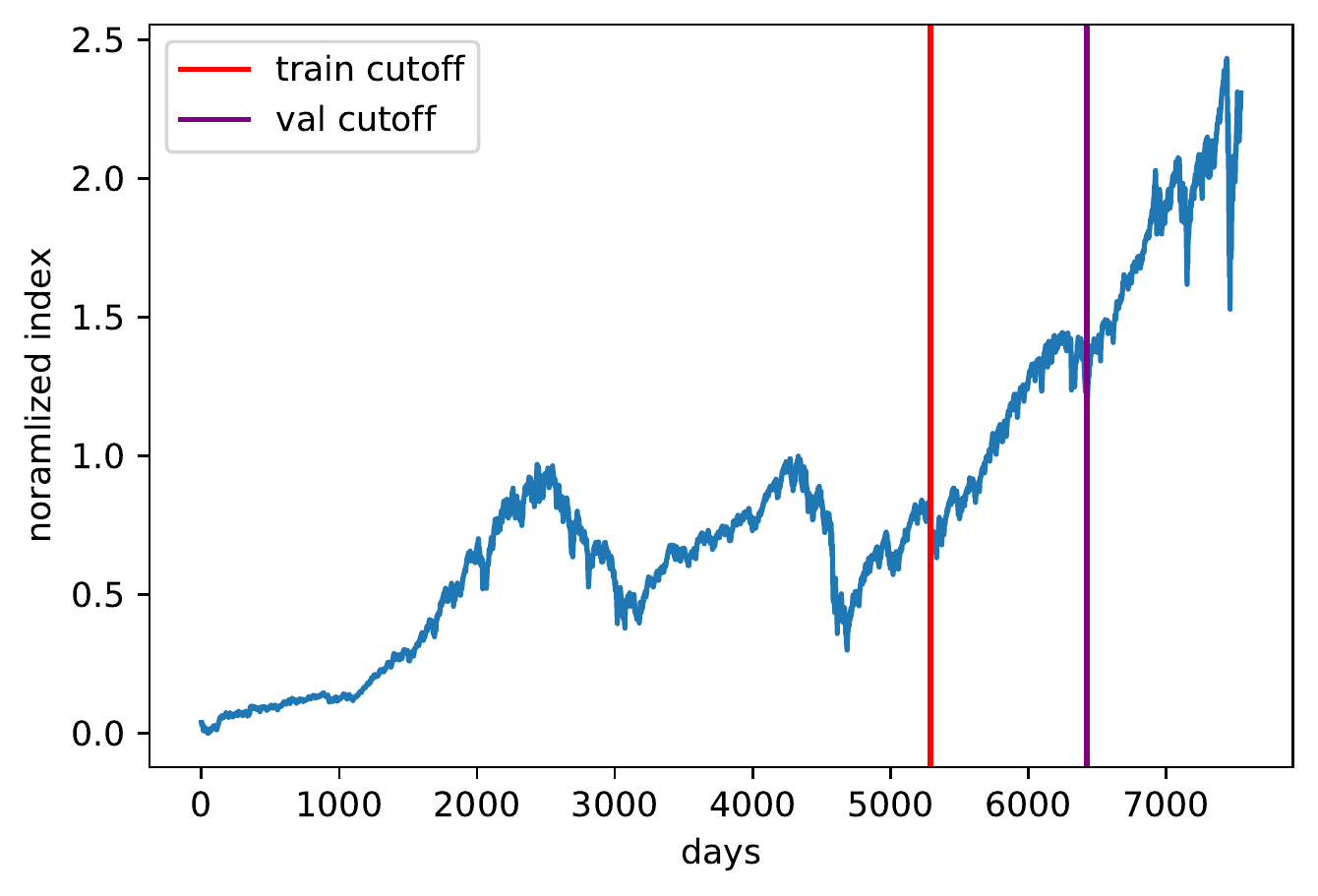}
\caption{Normalized values of the index with cutoffs for train and validation sets.}
\label{index_curve_r}
\end{figure}

The data is normalized to scale it between 0 and 1:
\begin{equation}
X = \frac{X-\mbox{min}(X_T)}{\mbox{max}(X_T)-\mbox{min}(X_T)},
\end{equation}
where $X_T$ is the training set. We use the training set in order to avoid leaking information about the validation and test sets into the training set. As a result, the values in the validation and test set are not strictly between 0 and 1. Similarly, we
normalize the values in target variable $y$. The graph of the normalized values of the target variable is given in Figure \ref{index_curve_r}. It is important to note that $y$ curve is increasing. The difference in train and validation set values may lead to underperformance.

\section{Results and analysis}
In this section, we use a fixed activation function ReLU in all models. We experiment with 3 different optimizers: SGD, RMRprop, and Adam. We use the default optimizer settings in Keras. The models are trained with batch size 32. During the initial training we apply Early Stopping callback based on the validation error to determine the optimal number of training epochs. We use the optimal number of training epochs to retrain the models on the combined train/validation set. After retraining we evaluate each model on the test set. The performance of the models is measured with mean average error (MAE) which is a standard accuracy criterion in machine learning.

\begin{figure}[h!]
\centering
\includegraphics[width=0.7\textwidth]{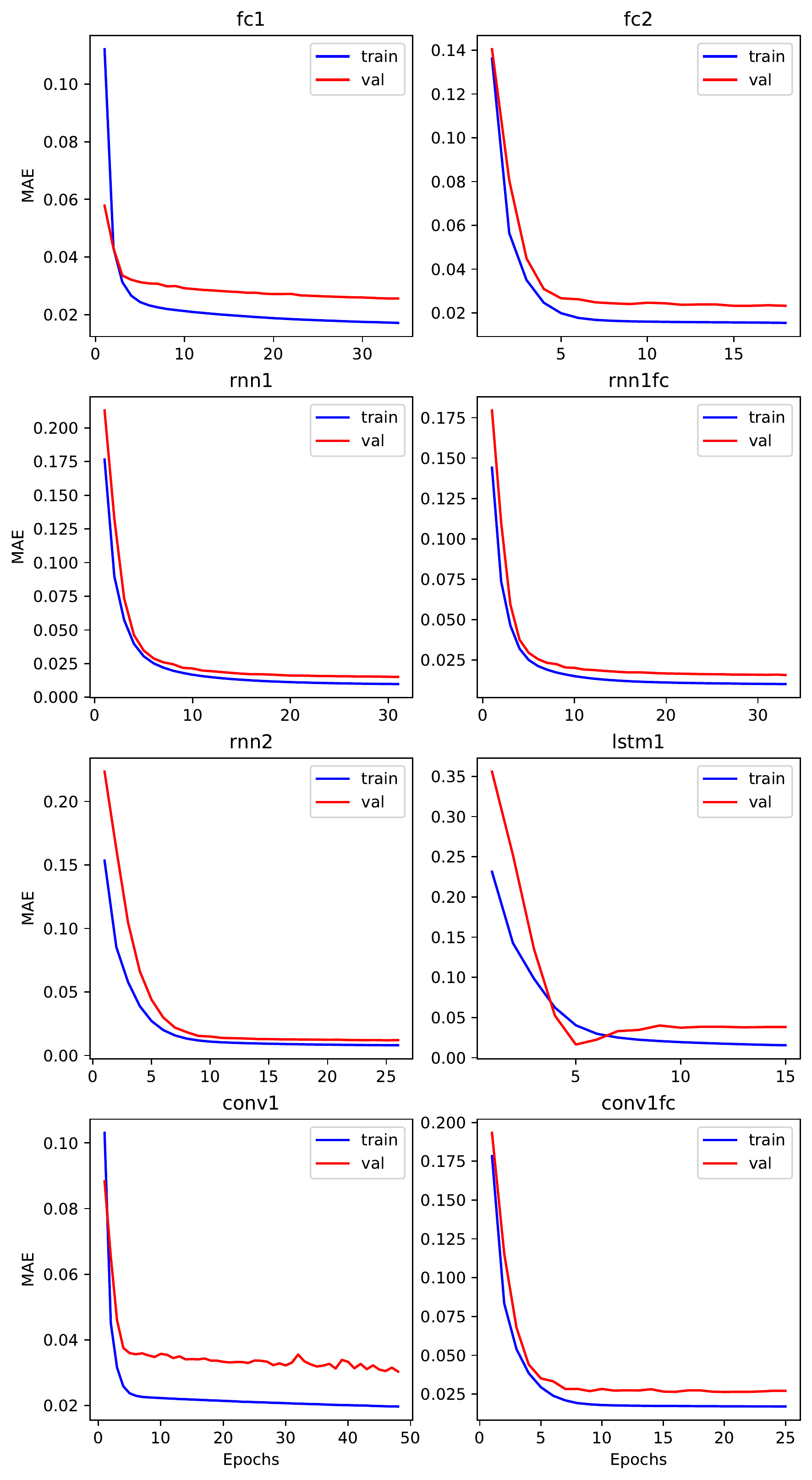}
\caption{Training results using SGD optimizer with batch size 32.}
\label{sgd32_ieee}
\end{figure}

The training results using the SGD optimizer are presented in Figure \ref{sgd32_ieee}. As can be seen in the figure  \texttt{rnn1, rnn1fc} and \texttt{rnn2} models appear to produce the best results.  All models reach their minimum error rates relatively fast. Additional training epochs appear to have little impact on model accuracy. The congruence of the training and validation curves indicates that the models are not overfitting. 
The minimum validation and test results for each model are summarized in Table \ref{tab:sgd}. We observe from the Table \ref{tab:sgd} that \texttt{rnn2} produces the lowest test error followed by \texttt{rnn1} and \texttt{rnn1fc}. 

\begin{table}[h!]
\centering
\caption{Summary of minimum validation and test MAE for each model based on SGD optimizer with batch size 32.  The \texttt{rnn2} model produces the lowest test error.}
\label{tab:sgd}
\begin{tabular}{lrr}
\toprule
Model &    Validation & Test   \\
\midrule
\texttt{fc1} &     0.0256 	&  0.0399\\
\texttt{fc2} &    0.0232 &  0.0422\\
\texttt{rnn1} &    0.0152  &  0.0207\\
\texttt{rnn1fc} &    0.0157  &  0.0214\\
\texttt{rnn2} &      0.0120  &  0.0186\\
\texttt{lstm1} &    0.0382  &  0.0736\\
\texttt{conv1} &    0.0304   &  0.0446\\
\texttt{conv1fc} &   0.0271  &  0.0408\\
\bottomrule
\end{tabular}
\end{table}

\begin{figure}[h!]
\centering
\includegraphics[width=0.7\textwidth]{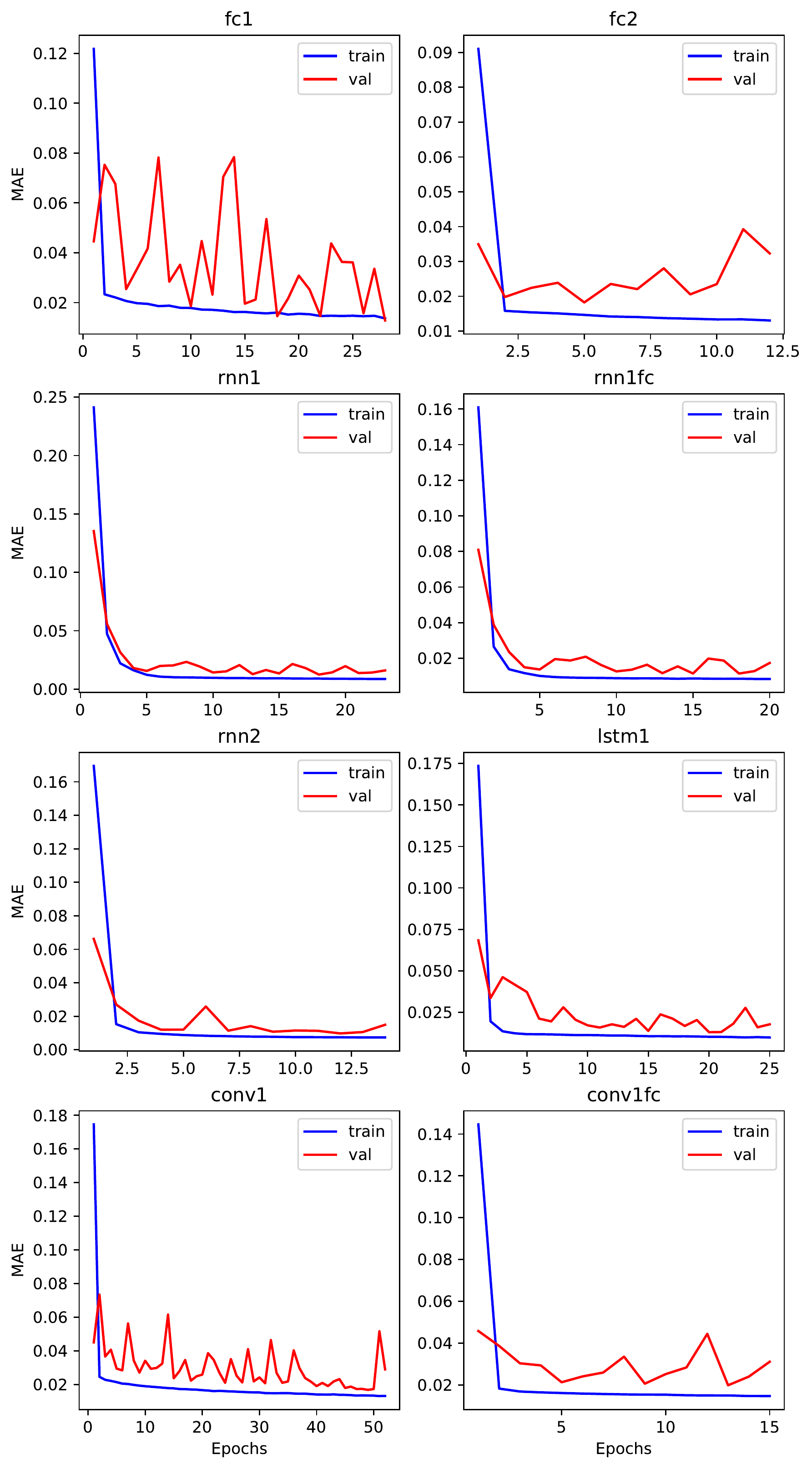}
\caption{Training results using RMSprop optimizer with batch size 32.}
\label{rmsprop32_ieee}
\end{figure}

The training results using the RMSprop optimizer are presented in Figure \ref{rmsprop32_ieee}. The validation curves appear to be much less stable compared to the SGD optimizer. As can be seen in the figure, \texttt{fc1} produces the lowest validation MAE followed by \texttt{rnn2} and \texttt{rnn1} models. The validation curves seem somewhat wilder with RMPprop optimizer compared to the SGD optimizer. It indicates a greater level of stochasticity in the gradient descent process. On other hand, the RMSprop algorithm leads to better results than SGD.
The minimum validation and test results for each model are summarized in Table \ref{tab:rms}. We observe from the Table \ref{tab:rms} that \texttt{rnn2} produces the lowest test error.  The error rate produced by the \texttt{rnn2} model is substantially lower than the rest of the models. We also note that the \texttt{rnn2} model achieves its minumum error rate in the smallest number of epochs.

\begin{table}[h!]
\centering
\caption{Summary of minimum validation and test MAE for each model based on RMS optimizer with batch size 32. The \texttt{rnn2} model produces the lowest test error among all the tested models. }
\label{tab:rms}
\begin{tabular}{lrr}
\toprule
Model &    Validation & Test   \\
\midrule
\texttt{fc1} &     0.0128 	&  0.0237 \\
\texttt{fc2} &    0.0322 &  0.0262 \\
\texttt{rnn1} &    0.0161  &  0.0202 \\
\texttt{rnn1fc} &    0.0174  &  0.0270 \\
\texttt{rnn2} &      0.0150  &  0.0148 \\
\texttt{lstm1} &    0.0178  &  0.0260  \\
\texttt{conv1} &    0.0291   &  0.0276 \\
\texttt{conv1fc} &   0.0311  &  0.0418  \\
\bottomrule
\end{tabular}
\end{table}

\begin{figure}[h!]
\centering
\includegraphics[width=0.7\textwidth]{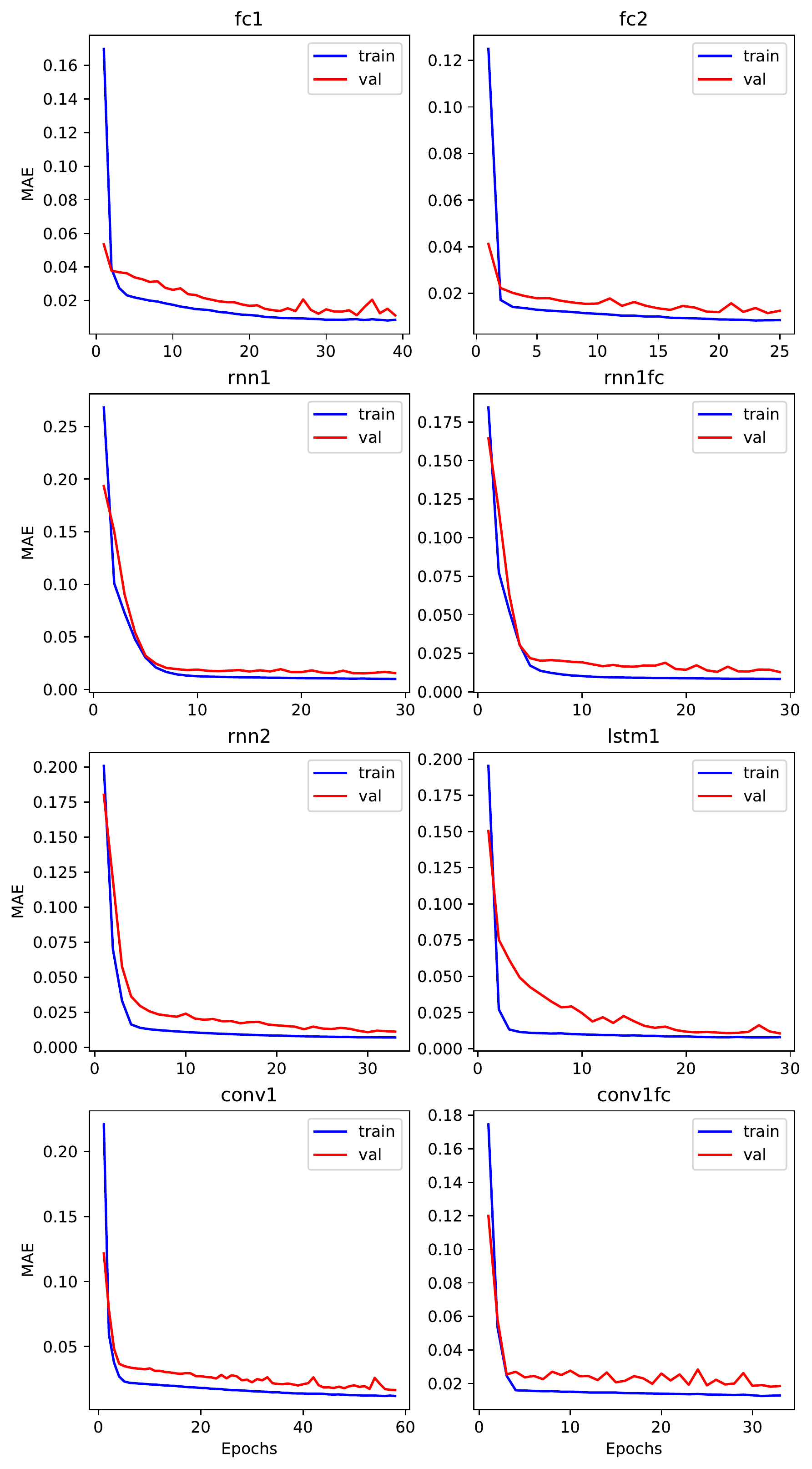}
\caption{Training results using Adam optimizer with batch size 32.}
\label{adam32_ieee}
\end{figure}

The training results using the Adam optimizer are presented in Figure \ref{adam32_ieee}. As can be seen in the figure, all models achieve low validation errors. The \texttt{lstm1} model produces the lowest validation MAE followed by \texttt{rnn2} and \texttt{fc1} models. The training and validation curves in Adam optimizer behave in similar fashion as the SGD optimizer. The training and validation curves reach a minimum relatively fast and further training yield little improvement. The training and validation curves are not diverging which indicates the absence of overfitting.
The minimum validation and test results for each model are summarized in Table \ref{tab:adam}. The test errors are noticeably greater than the validation errors which indicates the models overfit the data.

\begin{table}[h!]
\centering
\caption{Summary of minimum validation and test MAE for each model based on Adam optimizer with batch size 32. The \texttt{lstm1} model produces the lowest test error among all the tested models. }
\label{tab:adam}
\begin{tabular}{lrr}
\toprule
Model &    Validation & Test   \\
\midrule
\texttt{fc1} &     0.0113  	&  0.0198 \\
\texttt{fc2} &    0.0126  &  0.0324  \\
\texttt{rnn1} &    0.0156   &  0.0232  \\
\texttt{rnn1fc} &    0.0129   &  0.0201  \\
\texttt{rnn2} &      0.0112   &  0.0583  \\
\texttt{lstm1} &    0.0106   &  0.0666   \\
\texttt{conv1} &    0.0166    &  0.0284  \\
\texttt{conv1fc} &   0.0185   &  0.0278   \\
\bottomrule
\end{tabular}
\end{table}

\begin{figure}[h!]
\centering
\includegraphics[width=0.7\textwidth]{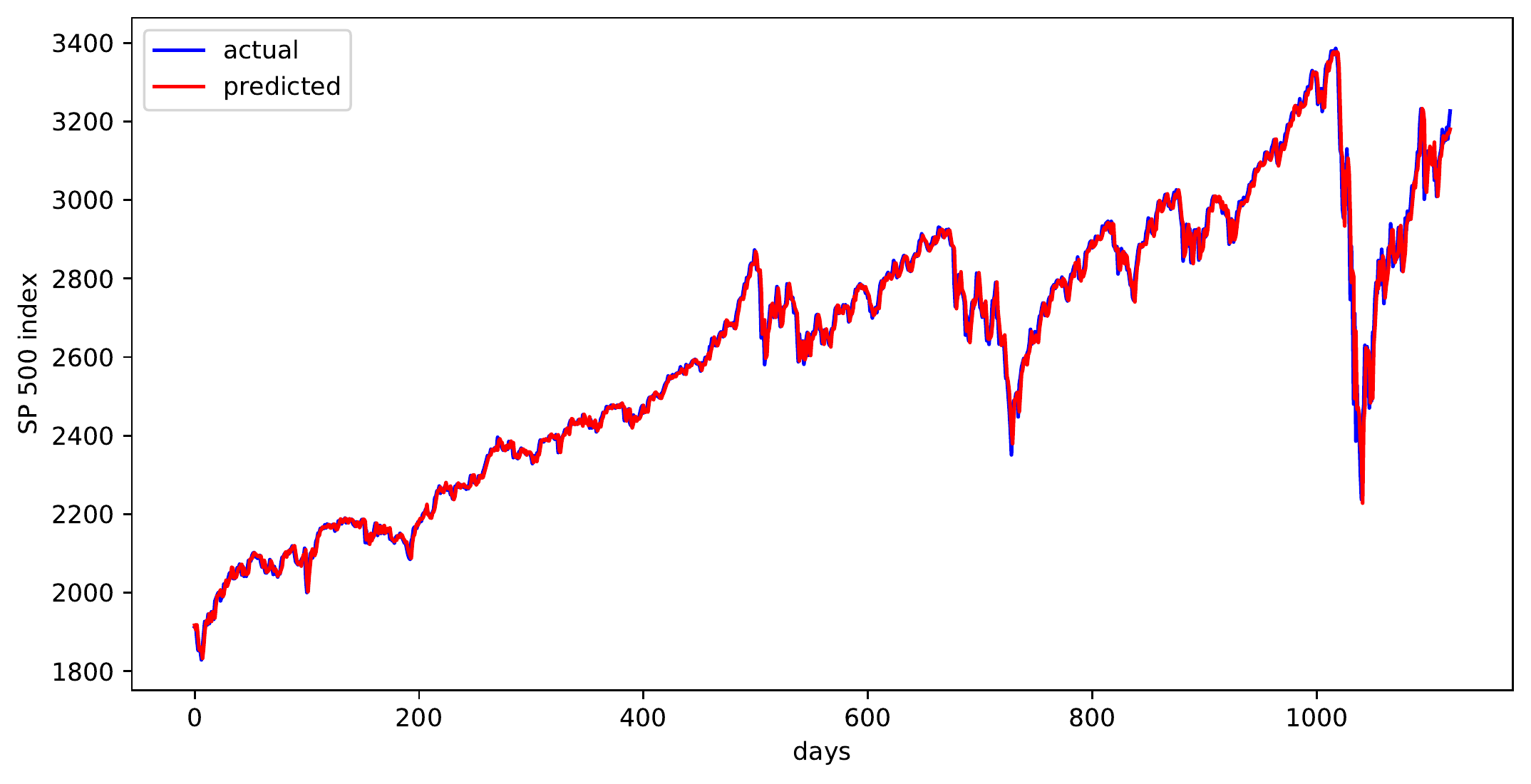}
\caption{The predicted and actual (normalized) values of the test set for S\&P 500 index using \texttt{rnn2} with RMSprop optimizer.}
\label{predicted_index_r}
\end{figure}

Stock price prediction is an inherently difficult task. Nevertheless, we have achieved a robust performance with the \texttt{rnn2} model using RMS optimizer. As shown in Figure \ref{predicted_index_r}, the predicted and actual values of the S\&P 500 index overlap almost entirely.
\section{Conclusion}
In this paper, we analyzed a variety of deep learning architectures for time-series forecasting. Concretely, we predicted the next day value of S\&P 500 index based on the information from the previous 14 trading days.
We compared eight different neural network models based on fully connected, convolutional, and recurrent layers. To expand our analysis we also applied three different optimizers. The numerical experiments reveal that a single layer recurrent neural network with RMSprop optimizer produces optimal results with validation and test MAE of 0.0150 and 0.0148 respectively.

The empirical results obtained in the paper demonstrate potential for the use of deep learning in market forecasting. Automated stock trading platforms based on deep learning may be valuable tools for brokerage firms involved in day trading. However, further analysis and experiments are necessary before deploying such models in production.
 We believe the analysis presented in this paper would be  helpful in stock price prediction using deep learning.

\end{document}